\documentclass[%
 reprint,
superscriptaddress,
 amsmath,amssymb,
 aps,
]{revtex4-2}

\usepackage{graphicx}
\usepackage{dcolumn}
\usepackage{bm}

\begin{document}

\preprint{APS/123-QED}

\title{Surface adsorbates suppress low-frequency noise for shallow nitrogen-vacancy centers}
\author{Zhiyang Yuan}
\affiliation{Department of Electrical and Computer Engineering, Princeton University, Princeton, New Jersey 08544, USA}

\author{David A. Fehr}
\affiliation{Department of Physics and Astronomy, University of Iowa, Iowa City, Iowa 52242, USA}%

\author{Kalliope Zervas}
\author{Sorawis Sangtawesin}
\author{Lila V. H. Rodgers}
\affiliation{Department of Electrical and Computer Engineering, Princeton University, Princeton, New Jersey 08544, USA}

\author{Patryk Gumann}
\affiliation{IBM T.J. Watson Research Center, Yorktown Heights, New York 10598, USA}

\author{Michael E. Flatt\'{e}}
\affiliation{Department of Physics and Astronomy, University of Iowa, Iowa City, Iowa 52242, USA}
\affiliation{Department of Applied Physics, Eindhoven University of Technology, Eindhoven 5612 AZ, The Netherlands}

\author{Nathalie P. de Leon}
\email{npdeleon@princeton.edu.}
\affiliation{Department of Electrical and Computer Engineering, Princeton University, Princeton, New Jersey 08544, USA}

\date{\today}

\begin{abstract}
Shallow nitrogen-vacancy (NV) centers in diamond are promising nanoscale quantum sensors, yet their coherence is strongly limited by surface-induced noise. Surface adsorbates are widely believed to be a major source of decoherence. Here, we test this assumption by characterizing shallow single NV centers under ultrahigh vacuum (UHV) conditions, where the diamond surface is kept free of adsorbates, and comparing their behavior to ambient conditions. Surprisingly, we observe a $\sim4\times$ reduction in the Hahn echo coherence time $T_2$ in UHV. By combining Hahn echo measurements in the single-quantum (SQ) and double-quantum (DQ) bases, we separate contributions from different noise sources and find that both electric and magnetic noise are enhanced in UHV. In contrast, $T_1$ measurements reveal an increased DQ $T_1$ in UHV, indicating suppressed electric field noise in the $\sim100$~MHz frequency regime. These results point to a modification of the surface noise spectrum upon adsorbate removal, with different frequency regimes arising from distinct microscopic mechanisms. Specifically, we find that the low frequency noise is consistent with increased surface charge in UHV that can be compensated by surface adsorbates in ambient conditions. Our findings highlight a complex and previously underappreciated role of surface adsorbates in shaping the noise environment of shallow NV centers, with important implications for nanoscale quantum sensing.
\end{abstract}

\maketitle

\section{\label{sec:intro}Introduction}
Nitrogen-vacancy (NV) centers in diamond have been explored as  powerful nanoscale quantum sensors, enabling the detection of magnetic, electric, and thermal signals with exceptional spatial resolution and sensitivity \cite{du2024single, maze2008nanoscale, balasubramanian2008nanoscale, maletinsky2012robust, taylor2008high}. Their unique combination of optical addressability and long spin coherence times under ambient conditions has enabled applications ranging from biological signal detection to studies of condensed-matter systems \cite{rodgers2024diamond,rovny2024nanoscale,casola2018probing,schirhagl2014nitrogen,kucsko2013nanometre, lovchinsky2016nuclear}. To further enhance sensitivity, it is essential to minimize the distance between the NV center and the target system, motivating the development of shallow NV centers located only a few nanometers beneath the diamond surface \cite{lovchinsky2016nuclear, mamin2013nanoscale, Sangtawesin2019a,rosskopf2014investigation, romach2015spectroscopy, myers2014probing}.

However, bringing NV centers closer to the surface introduces significant challenges. Shallow NV centers typically exhibit reduced spin coherence times ($T_2$) \cite{Sangtawesin2019a, myers2014probing, ohno2012engineering, ofori2012spin} and increased charge state instability \cite{bluvstein2019identifying, yuan2020charge, yamano2017charge, giri2023charge, haruyama2023charge}, both of which degrade sensing performance. These effects are widely attributed to surface-related noise, yet the microscopic origins of this noise remain incompletely understood. Identifying and disentangling dominant noise sources is therefore essential for realizing the full potential of shallow NV-based quantum sensing applications.

Several mechanisms have been proposed to contribute to decoherence and instability in shallow NV centers. Electric field noise arising from fluctuating surface charges has been shown to play an important role \cite{kim2015decoherence}. In addition, structural defects at or near the diamond surface---including $\mathrm{sp}^2$-bonded carbon and other surface states---can introduce deep electronic trap states that contribute to noise \cite{stacey2019evidence}. Furthermore, subsurface damage resulting from ion implantation, such as multivacancy complexes, can act as sources of paramagnetic noise and degrade the coherence of nearby NV centers \cite{favaro2017tailoring, yamamoto2013extending}.

In parallel, the diamond surface itself is a complex and dynamic environment. Under ambient conditions, surfaces are typically covered by adsorbates, including adventitious carbon, water, and other contaminants whose composition depends sensitively on the surrounding atmosphere \cite{barr1995nature, greczynski2020x, grey2024defining}. These adsorbates, along with variations in surface termination, are known to influence surface conductivity and electronic structure \cite{rivero2016surface, chakrapani2005studies, nebel2006surface}. Despite extensive study, their specific contributions to the noise of shallow NV centers remain difficult to isolate due to the coexistence of multiple competing effects.

Ultra-high vacuum (UHV) provides a controlled environment to systematically investigate these surface-related phenomena by removing adsorbates and enabling reproducible surface preparation \cite{yuan2026integratedultrahighvacuumcluster}. By studying NV centers in a well-defined surface environment, UHV experiments offer a unique opportunity to disentangle the relative contributions of adsorbates, surface defects, and subsurface damage to NV decoherence and charge instability. Furthermore, such studies are essential for the engineering of diamond surfaces, particularly for applications involving hydrogen-terminated diamond and other functionalized interfaces \cite{kaviani2014proper}.

In this work, we directly compare the properties of the same individual shallow NV centers under ambient and UHV conditions. By performing longitudinal ($T_1$) and transverse ($T_2$) relaxation measurements, we extract detailed information about the underlying noise environment and its dependence on surface conditions. Specifically, we measure NV spin dynamics in both the single-quantum (SQ) and double-quantum (DQ) bases to disentangle electric and magnetic noise contributions, analyze the depth dependence of noise rates to identify microscopic origins, and reconstruct the noise spectrum through dynamical decoupling measurements. This approach allows us to isolate the role of surface adsorbates and probe the dynamics of the vacuum–diamond interface. Our results provide new insights into the mechanisms limiting shallow NV performance. 

\section{Comparison of NV Centers in Air and in UHV\label{sec:air vs UHV}}

To investigate the influence of surface conditions on shallow NV centers, we directly compare their properties under ambient air and UHV conditions. The diamond sample is loaded into a custom-built UHV chamber (see the design of the chamber in Ref.~\cite{yuan2026integratedultrahighvacuumcluster}), where it undergoes an in situ anneal at $350\,^{\circ}\mathrm{C}$ to remove surface adsorbates (Fig.~\ref{fig:shorter_T2}(a)). Details of diamond sample preparation and the UHV annealing procedure are provided in Appendix~\ref{sec:sample preparation}. During measurements, the chamber pressure is maintained at approximately $5\times10^{-10}$~mbar, with the residual gas composition dominated by hydrogen. Under these conditions, the diamond surface is expected to remain largely free of adsorbates throughout the experiment, as confirmed by the experiments presented in Ref.~\cite{yuan2026integratedultrahighvacuumcluster} and Appendix~\ref{sec:long term measurements}. To further verify this, we performed NV NMR measurements targeting the proton signal from surface adsorbates and observed no detectable proton signal under UHV conditions. Details of these measurements, together with an estimate of the surface proton density, are provided in Appendix~\ref{sec:proton NMR}.

One key aspect of this study is the one-to-one comparison of the same individual NV centers measured in air and in UHV. NV locations are tracked across different environments (see Appendix~\ref{sec:air UHV confocals} for details), enabling a direct assessment of how the surface environment impacts their properties. Figure~\ref{fig:shorter_T2}(b) shows representative Hahn echo measurements of a shallow NV center in air and in UHV. Counterintuitively, we observe a pronounced reduction in the coherence time $T_2$ after the removal of surface adsorbates. Across multiple NV centers, the ratio $T_{2, \mathrm{air}} / T_{2, \mathrm{UHV}}$ ranges from 3.0 to 5.4, with larger reductions observed for shallower NV centers (Fig.~\ref{fig:shorter_T2}(c)). This depth dependence strongly suggests that the effect originates from the diamond surface. To exclude alternative experimental artifacts for the reduced $T_2$ in UHV---such as differences in optical excitation conditions or potential influences from vacuum instrumentation (e.g., ion pump and ion gauge)---we performed a series of control experiments; details are provided in Appendix~\ref{sec:t2 deep nv}--\ref{sec:ion pump ion gauge}. 

To test whether this degradation arises from irreversible surface modification in UHV, we unload the sample from the chamber and remeasure the same NV centers under ambient conditions (Fig.~\ref{fig:shorter_T2}(d)). In all cases, the longer $T_2$ times are restored upon exposure to air. Repeating the UHV--air cycling process a few times yields consistent and reproducible changes in $T_2$. This reversibility indicates that the observed coherence degradation in UHV is not due to permanent damage to the diamond surface, but instead arises from a dynamic and environment-dependent surface condition.

\begin{figure}
    \centering
    \includegraphics{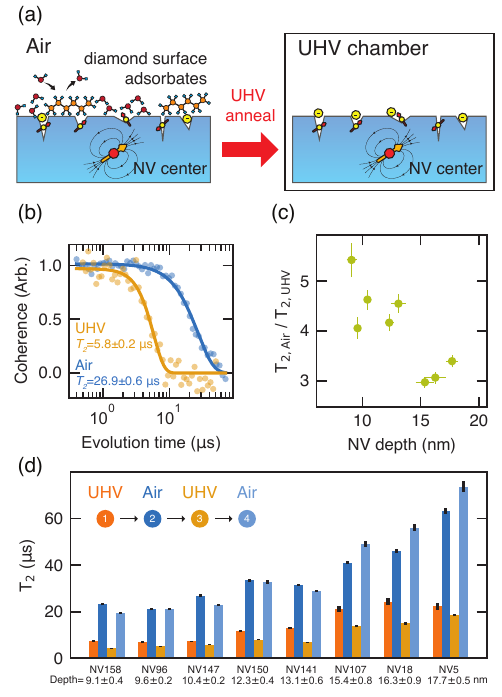}
    \caption{
    (a) Schematic of a near-surface nitrogen-vacancy (NV) center in diamond. Under ambient (air) conditions, the diamond surface is covered by adsorbates such as water and hydrocarbons, whereas under ultra-high vacuum (UHV) conditions, these adsorbates are removed via low-temperature annealing, leaving a clean surface.
    (b) Hahn echo coherence time $T_2$ measured for the same NV center in air (blue) and in UHV (orange), showing a $4.6\times$ longer $T_2$ in air. The bias magnetic field is 420~G and is aligned with the NV axis.
    (c) Ratio of $T_2$ measured in air to that in UHV for eight NV centers, plotted as a function of NV depth. The depths are independently calibrated via proton NMR measurements in immersion oil \cite{pham2016nmr}. Higher ratios are observed for NV centers closer to the surface.
    (d) Histogram of $T_2$ values for the same eight NV centers measured sequentially in UHV and in air, demonstrating reproducible shortening of $T_2$ under UHV conditions and revival of longer $T_2$ in air. The diamond sample is cleaned using a piranha solution between each measurement step. Step 4 is measured in air directly after unloading the sample from the UHV chamber without any cleaning.}
    \label{fig:shorter_T2}
\end{figure}

\section{Electric and Magnetic Contributions to $T_2$ decoherence\label{sec:E_B_T2}}

To identify the mechanism underlying the increased decoherence in UHV, we separate the contributions of electric and magnetic noise to the NV spin dynamics. Previous studies have shown that electric field noise can play a significant role for shallow NV centers, and that screening by high-dielectric-constant liquids can substantially extend coherence times \cite{kim2015decoherence}. This raises the possibility that the removal of surface adsorbates in UHV reduces dielectric screening and thereby enhances electric field noise.

\begin{figure*}
    \centering
    \includegraphics{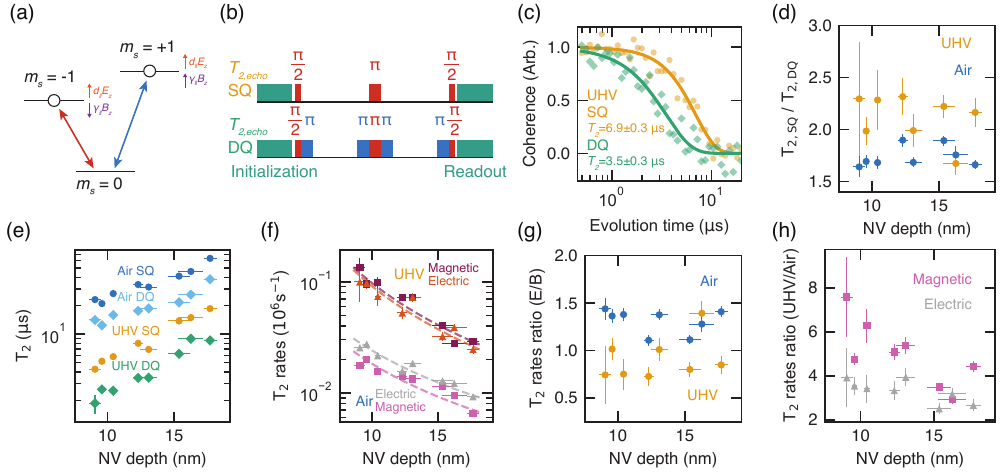}
    \caption{
    (a) Energy level diagram of the ground-state electronic spin of the negatively charged NV center, indicating the two microwave-driven transitions. Magnetic field noise along the NV axis shifts the $m_s=+1$ and $m_s=-1$ levels in opposite directions, whereas electric field noise shifts them in the same direction. 
    (b) Pulse sequences for Hahn echo coherence measurements in the single-quantum (SQ) and double-quantum (DQ) bases. In the SQ protocol, the coherence of the superposition between $m_s=0$ and $m_s=-1$ is probed. In the DQ protocol, a superposition between $m_s=+1$ and $m_s=-1$ is generated using two microwave tones, as the direct transition between these states is forbidden. Red microwave pulses are at the transition frequency between $m_s=0$ and $m_s=-1$, while blue pulses are applied at the $m_s=0$ to $m_s=+1$ frequency. Spin initialization and readout are performed using green (561 nm) laser pulses.
    (c) Representative SQ and DQ Hahn echo measurements for a single NV center under UHV conditions, showing that the SQ coherence time $T_2$ is approximately twice as long as the DQ $T_2$. 
    (d) Ratio of SQ to DQ $T_2$ for eight NV centers measured in UHV and in air, plotted as a function of NV depth. The ratio is generally higher under UHV conditions. 
    (e) Comparison of SQ and DQ $T_2$ as a function of NV depth in UHV and in air. Both SQ and DQ coherence times are consistently longer in air, with shorter $T_2$ observed for shallower NV centers. 
    (f) Decomposed magnetic and electric contributions to the decoherence rates in UHV and in air, plotted versus NV depth. The data are fitted to power-law functions (dashed lines), yielding exponents of $-2.13 \pm 0.21$ (UHV, magnetic), $-2.19 \pm 0.27$ (UHV, electric), $-1.67 \pm 0.19$ (air, magnetic), and $-1.63 \pm 0.14$ (air, electric).
    (g) Ratio of electric and magnetic decoherence rates. In air, electric decoherence rates are higher than magnetic decoherence rates, while in UHV electric contributions to the noise are slightly lower than magnetic contributions. 
    (h) Ratio of UHV and air decoherence rates. Both magnetic and electric decoherence rates are higher in UHV. Magnetic decoherence rates are enhanced more in UHV and show more obvious NV depth dependence compared to the electric decoherence rates.}
    \label{fig:DQ_T2}
\end{figure*}

To test this hypothesis, we perform Hahn echo measurements in both the SQ and DQ bases (Fig.~\ref{fig:DQ_T2}(a)) \cite{kim2015decoherence, PhysRevLett.113.030803}. While direct driving between the $m_s=-1$ and $m_s=+1$ states is forbidden, the DQ sequence is implemented by introducing additional $\pi$ pulses via the $m_s=0 \leftrightarrow +1$ transition (Fig.~\ref{fig:DQ_T2}(b)). Figure~\ref{fig:DQ_T2}(c) shows a representative comparison of SQ and DQ Hahn echo measurements for an NV center in UHV.

The DQ basis provides enhanced sensitivity to magnetic noise while suppressing sensitivity to electric field fluctuations along the NV axis. For the NV ground-state spin states, the energy splitting between $m_s=-1$ and $m_s=+1$ is twice as sensitive to magnetic field fluctuations as the $m_s=0 \leftrightarrow -1$ transition. At the same time, longitudinal electric field noise ($E_z$) shifts the $m_s=\pm1$ levels equally in the same direction and is therefore rejected in the DQ protocol (Fig.~\ref{fig:DQ_T2}(a)) \cite{dolde2011electric}. Transverse electric field noise is further suppressed by applying a strong bias magnetic field along the NV axis \cite{kim2015decoherence}. In the limit of purely magnetic noise, one therefore expects $T_{2,\mathrm{SQ}} = 4T_{2,\mathrm{DQ}}$, as the decoherence measurement is sensitive to the variance of magnetic noise. Experimentally, we observe $T_{2,\mathrm{SQ}}/T_{2,\mathrm{DQ}}$ ratios between 1.5 and 2.5 (Fig.~\ref{fig:DQ_T2}(d)), indicating a significant contribution from electric field noise. Notably, this ratio increases in UHV compared to ambient conditions, suggesting a change in the relative balance between electric and magnetic noise sources.

To quantify these contributions, we decompose the total SQ decoherence rate into electric and magnetic components:
\begin{equation}
\label{eq:T2_SQ}
    \frac{1}{T_{2, \mathrm{SQ}}} = \frac{1}{T_{2, \mathrm{B}}} + \frac{1}{T_{2, \mathrm{E}}}.
\end{equation}
From the analysis above, the DQ decoherence is more sensitive to the magnetic component while insensitive to the electric component:
\begin{equation}
\label{eq:T2_DQ}
    \frac{1}{T_{2, \mathrm{DQ}}} = \frac{4}{T_{2, \mathrm{B}}}.
\end{equation}
Therefore, we can quantify the electric and magnetic contributions to the NV decoherence by combining SQ and DQ $T_2$ measurements:
\begin{equation}
\label{eq:T2_decomposition}
\begin{aligned}
    \frac{1}{T_{2, \mathrm{B}}} &= \frac{1}{4T_{2, \mathrm{DQ}}},\\
    \frac{1}{T_{2, \mathrm{E}}} &= \frac{1}{T_{2, \mathrm{SQ}}} - \frac{1}{4T_{2, \mathrm{DQ}}}.
\end{aligned}
\end{equation}
A detailed discussion of this decomposition, following the formalism of Ref.~\cite{candido2024interplay}, is provided in Appendix~\ref{sec:decoherence rates decomposition}.

Figure~\ref{fig:DQ_T2}(e) shows the depth dependence of $T_2$ measured in both SQ and DQ bases under air and UHV conditions. In all cases, $T_2$ decreases for shallower NV centers, consistent with a surface-originated noise source. Moreover, both SQ and DQ $T_2$ times are systematically shorter in UHV.
From the combined $T_2$ times, we extract the effective electric and magnetic decoherence rates for each NV center (Fig.~\ref{fig:DQ_T2}(f)). We find that both contributions are enhanced in UHV (Fig.~\ref{fig:DQ_T2}(h)). Under ambient conditions, electric noise is higher than magnetic noise, whereas in UHV the magnetic contribution becomes comparable to, and in some cases exceeds, the electric contribution (Fig.~\ref{fig:DQ_T2}(g)). These results demonstrate that the increased decoherence in UHV cannot be explained solely by the loss of dielectric screening of electric field noise. Instead, both electric and magnetic noise are amplified, with a stronger relative increase in the magnetic component.

Furthermore, we analyze the depth dependence of the extracted noise components shown in Fig.~\ref{eq:T2_DQ}(f). Comparing our results with existing models for shallow NV decoherence \cite{candido2024interplay}, we find that the depth dependence of both electric and magnetic noise follows a scaling closer to $1/d_{\mathrm{NV}}^{2}$ rather than $1/d_{\mathrm{NV}}^{4}$, where $d_{\mathrm{NV}}$ denotes the NV depth. This scaling suggests that the electric noise is dominated by point-like fluctuating surface charges rather than dipolar fluctuations, whereas the magnetic noise is more consistent with mobile charged species at the diamond surface than with fluctuating surface spins. A detailed analysis of the NV depth dependence of surface noise and a modified model accounting for the deviation of the depth-scaling exponent are presented in Appendix~\ref{sec:depth scaling model}. 

\section{NV noise spectrum\label{sec:noise_spectrum}}
\begin{figure}
    \centering
    \includegraphics{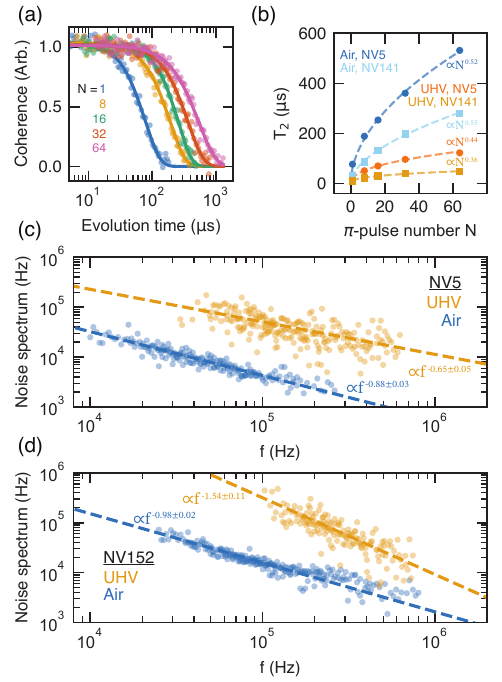}
    \caption{
    (a) Dynamical decoupling measurements of NV coherence using sequences with varying numbers of $\pi$ pulses, performed on NV5 in air under an applied magnetic field of 420~G. 
    (b) Extracted $T_2$ coherence times as a function of the number of $\pi$ pulses, demonstrating the extension of coherence with increasing $\pi$ pulse numbers. Example data from two NV centers measured in air and in UHV are shown. Dashed lines are fits to power-law scaling, yielding exponents of $0.52 \pm 0.02$ (air, NV5), $0.55 \pm 0.02$ (air, NV141), $0.44 \pm 0.02$ (UHV, NV5), and $0.36 \pm 0.01$ (UHV, NV141). 
    (c,d) Noise spectra obtained via spectral decomposition of the dynamical decoupling data in air and in UHV for NV5 (c) and NV152 (d). In all cases, the spectra exhibit a clear power-law dependence over the measured frequency range. Dashed lines indicate fits to a power-law model, yielding exponents of $-0.65 \pm 0.05$ (UHV, NV5), $-0.88 \pm 0.03$ (air, NV5), $-1.54 \pm 0.11$ (UHV, NV152), and $-0.98 \pm 0.02$ (air, NV152). 
    }
    \label{fig:DD_spec}
\end{figure}

To investigate changes in the spectral density of the noise bath under UHV conditions, we perform dynamical decoupling measurements on shallow NV centers. Dynamical decoupling sequences are widely used to extend qubit coherence while selectively filtering environmental noise at different frequencies~\cite{romach2015spectroscopy,myers2017double,bar2012suppression,Sangtawesin2019a}. An example dataset is shown in Fig.~\ref{fig:DD_spec}(a). As the number of applied $\pi$ pulses increases, the extracted $T_2$ coherence times exhibit a power-law scaling with the number of $\pi$ pulses, $T_2 \propto N^{s}$, as shown in Fig.~\ref{fig:DD_spec}(b). We measure exponents $s$ in the range of $0.36$ to $0.55$ for both air and UHV conditions, which deviate from the expected value of $2/3$ for a slowly fluctuating spin bath \cite{de2010universal}. Notably, for the same NV center, the difference in the fitted exponent $s$ between air and UHV measurements suggests a modification of the noise bath dynamics, potentially reflecting a change in the correlation time of the dominant noise sources. This distinction is further elucidated using the spectral decomposition technique discussed below.

At large numbers of $\pi$ pulses, dynamical decoupling sequences act as narrow-band filter functions that selectively couple the NV spin to environmental noise at frequencies determined by the inverse of the interpulse spacing. By sweeping the evolution time, we reconstruct the noise spectral density over a wide frequency range. Spectra obtained from sequences with $N=8, 16, 32,$ and $64$ pulses are combined to extend the accessible frequency window. 

Figures~\ref{fig:DD_spec}(c,d) show representative noise spectra measured in air and in UHV. Over the frequency range from $\sim$10~kHz to 800~kHz, the extracted spectra exhibit a clear power-law dependence on frequency. Within this range, the noise spectral density in UHV is consistently higher than that measured in air. The extracted power-law exponents lie between 0.6 and 1.5, deviating from the $1/f^2$ scaling expected for a single Lorentzian spectrum in the high-frequency limit. Such behavior can be understood by considering a distribution of Lorentzian contributions arising from multiple surface trap centers, each characterized by a distinct correlation time. Different distributions of these fluctuators naturally give rise to power-law spectra with varying exponents, while variations between NV centers likely reflect differences in their local environments. Notably, we also observe a change in the power-law exponent between air and UHV for the same NV center, indicating a redistribution of the noise spectral weight across frequencies. However, the direction of this change is not consistent: for NV5, the exponent increases from $-0.88$ (air) to $-0.65$ (UHV), whereas for NV152 it decreases from $-0.98$ (air) to $-1.54$ (UHV). This contrasting behavior suggests that the difference between the air and UHV $T_2$ values and noise spectra cannot be explained by a simple shift in fluctuator correlation times alone, which would be expected to produce a consistent change in the spectral exponent when switching from air to UHV. Instead, it points to an overall increase in noise density within the $\sim$10~kHz to 800~kHz frequency range, accompanied by a modification of the underlying fluctuator distribution. A detailed discussion of the origin of the observed $1/f^{\alpha}$ spectra, based on an ensemble of Lorentzian fluctuators with a distribution of correlation times, is presented in Appendix~\ref{sec:1/f spec}. Using this model, we further estimate the corresponding increase in surface charge density under UHV conditions.

\section{$T_1$ Relaxation and Higher-Frequency Noise}

To further probe the noise environment, we measure the longitudinal relaxation time $T_1$ of shallow NV centers under ambient and UHV conditions. Unlike $T_2$, which is sensitive to relatively low-frequency noise, $T_1$ probes noise at the transition frequencies of the NV spin levels, typically in the $\sim100$~MHz to GHz range, thereby providing complementary spectral information.

\begin{figure}
    \centering
    \includegraphics{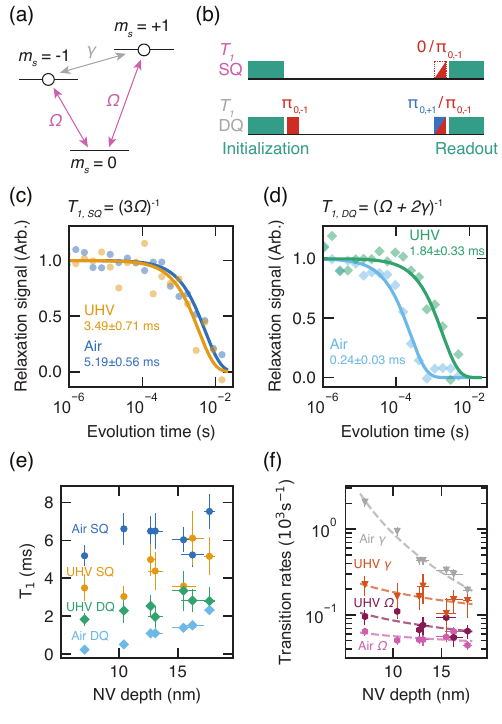}
    \caption{
    (a) Energy level diagram of the NV center spin states, indicating SQ and DQ relaxation transitions, with corresponding rates $\Omega$ and $\gamma$. 
    (b) Pulse sequences for SQ and DQ relaxation measurements. In the SQ (DQ) protocol, the NV spin is initialized in the $m_s=0$ ($m_s=-1$) state, and the relaxation to the $m_s=-1$ ($m_s=+1$) state is measured. 
    (c,d) Comparison of SQ $T_1$ (c) and DQ $T_1$ (d) in UHV and in air for the same NV center; data (points) are fitted to exponential functions (lines). These measurements are performed at a magnetic field of 18~G.
    (e) SQ and DQ $T_1$ as a function of NV depth in UHV and in air, showing shorter SQ $T_1$ but longer DQ $T_1$ in UHV. 
    (f) Decomposed SQ ($\Omega$) and DQ ($\gamma$) relaxation rates in UHV and in air versus NV depth. Dashed lines indicate power-law fits, yielding exponents of $-0.50 \pm 0.28$ (UHV, $\Omega$), $-0.55 \pm 0.19$ (UHV, $\gamma$), $-0.22 \pm 0.16$ (air, $\Omega$), and $-2.57 \pm 0.16$ (air, $\gamma$).}
    \label{fig:T1}
\end{figure}

The $T_1$ relaxation can be measured in both the SQ and DQ bases (Fig.~\ref{fig:T1}(a)), with corresponding pulse sequences shown in Fig.~\ref{fig:T1}(b). These measurements probe distinct noise channels. SQ transitions ($m_s=0 \leftrightarrow \pm1$) are sensitive to both magnetic and electric noise, whereas the DQ transition ($m_s=-1 \leftrightarrow +1$) is magnetic-dipole forbidden and is therefore predominantly sensitive to electric field noise \cite{lin2022diamond}.

Figures~\ref{fig:T1}(c,d) show representative comparisons of $T_1$ measurements in air and in UHV for both SQ and DQ bases. The measurements are performed at a low magnetic field of $18$~G, corresponding to transition frequencies of $\sim 2.82$~GHz for the SQ transition and $\sim 100$~MHz for the DQ transition. The depth dependence of the extracted $T_1$ times is summarized in Fig.~\ref{fig:T1}(e). In contrast to the $T_2$ behavior, we observe a slight decrease in SQ $T_1$ and an increase in DQ $T_1$ under UHV conditions.

To quantify these trends, we express the relaxation in terms of effective rates associated with different noise channels. Following Ref.~\cite{myers2017double}, we have
\begin{equation}
\label{eq:T1_SQ DQ}
    T_{1, \mathrm{SQ}} = (3\Omega)^{-1}, \qquad T_{1, \mathrm{DQ}} = (\Omega + 2\gamma)^{-1}.
\end{equation}
Therefore, we can calculate $\Omega$ and $\gamma$ relaxation rates by combining the $T_{1, \mathrm{SQ}}$ and $T_{1, \mathrm{DQ}}$ times for each NV center. The SQ relaxation rate $\Omega$ captures noise at the GHz frequency scale, while the DQ relaxation rate $\gamma$ reflects electric field noise near $100$~MHz. The NV depth dependence of these rates is shown in Fig.~\ref{fig:T1}(f). We observe a pronounced reduction in $\gamma$ in UHV, indicating a suppression of electric field noise at $\sim 100$~MHz. The remaining $\gamma$ rates in UHV are close to the bulk limit of approximately $0.14 \times 10^{3}$~s$^{-1}$ (see Appendix~\ref{sec:extra_T1}), suggesting the removal of a surface electric noise channel at $\sim 100$~MHz in UHV. In contrast, $\Omega$ is slightly increased in UHV, suggesting a modest enhancement of high-frequency magnetic noise in the GHz regime.

We also note that the depth dependence of $\Omega$ and $\gamma$ (Fig.~\ref{fig:T1}(f)) differs from that of the $T_2$ decoherence rates (Fig.~\ref{fig:DQ_T2}(f)). At these higher frequencies, surface-induced noise decays rapidly with distance, and bulk contributions begin to dominate the relaxation processes at NV depths investigated here. A more pronounced $T_1$ depth dependence is observed for NV centers shallower than 10~nm and the depth scaling is more consistent with fluctuating surface dipoles (see additional measurements and analysis of $T_1$ depth scaling in air and in an oil-immersion confocal microscope in Appendix~\ref{sec:extra_T1}). The distinct depth-scaling exponents may further indicate different microscopic origins of the noise processes governing $T_1$ and $T_2$.

\section{Discussion\label{sec:discussion}}

The observation that removing surface adsorbates under UHV leads to reduced, rather than improved, coherence of shallow NV centers challenges the common assumption that surface adsorbates are a primary source of decoherence and that cleaner surfaces should always enhance NV performance.

By combining $T_2$ coherence measurements, dynamical decoupling noise spectroscopy, and $T_1$ relaxation measurements, we identify a frequency-dependent restructuring of the surface noise environment under UHV conditions. At low frequencies probed by $T_2$ measurements, both electric and magnetic noise contributions increase after removing surface adsorbates. The depth dependence of the electric noise component is consistent with fluctuating surface charges rather than fluctuating surface dipoles. Similarly, the magnetic noise scaling is more consistent with magnetic fields generated by the motion of surface charges than with surface spin fluctuations. The reconstructed noise spectra show an overall enhancement of noise in UHV across the low-frequency regime ($\sim 10$~kHz to $1$~MHz), which can be explained by an increased density of fluctuating surface charges after adsorbate removal. Additional evidence for UHV-induced surface charging is provided by enhanced surface PL observed both inside the UHV environment (see Ref.~\cite{yuan2026integratedultrahighvacuumcluster}) and after removing the sample from UHV (see Appendix~\ref{sec:air after UHV}). These observations indicate that the NV center itself, under green laser excitation, can contribute to the generation of excess surface charge when dissipation from atmospheric adsorbates is absent. This highlights an intrinsic challenge for studying and utilizing shallow NV centers under UHV conditions.

In contrast, noise contributions at higher frequencies exhibit a different response to adsorbate removal. DQ $T_1$ measurements show that electric field noise at intermediate frequencies ($\sim100$~MHz) is reduced under UHV conditions. The depth dependence of this contribution differs from the low-frequency electric noise extracted from $T_2$ measurements, suggesting that distinct microscopic mechanisms dominate in these two frequency regimes. The intermediate-frequency noise is more consistent with fluctuating surface dipoles associated with adsorbates, and its removal after UHV annealing can explain the observed increase in DQ $T_1$ under UHV conditions.

At even higher frequencies in the GHz range, we observe a slight increase in SQ $T_1$ relaxation rates under UHV. However, the investigated NV centers exhibit relaxation times close to the expected bulk-limited regime, where surface contributions become relatively small compared with bulk relaxation processes. As a result, the influence of UHV-induced surface modifications is difficult to isolate in this frequency range. Future low-temperature measurements, where bulk relaxation rates are suppressed, may enable a more sensitive investigation of high-frequency surface noise contributions.

By combining the noise spectra extracted from $T_2$ and $T_1$ measurements (Appendix~\ref{sec:combined spectrum}), we find that surface-induced decoherence cannot be described by a single universal noise mechanism. Instead, different microscopic processes dominate at different frequencies: fluctuating surface charges govern low-frequency decoherence, adsorbate-related dipole fluctuations contribute at intermediate frequencies, and bulk processes become increasingly important at higher frequencies. 

Beyond spin coherence and relaxation, the removal of adsorbates also strongly influences the charge-state stability of shallow NV centers. We observe reduced charge-state stability for NV centers shallower than 6~nm under UHV conditions (see Appendix~\ref{sec:air UHV confocals}). Ensemble measurements also show a reduced $\mathrm{NV}^{-}$ population after green laser excitation in UHV, whereas ambient measurements result in a higher $\mathrm{NV}^{-}$ population after illumination (see Appendix~\ref{sec:ensemble measurements}). These observations suggest that surface adsorbates and their associated electrostatic environment play an important role in stabilizing the NV charge environment as well.

\section{\label{sec:conclusion}Conclusion}
In this work, we present the first detailed study of NV center properties in a UHV environment and under a pristine surface. Systematic characterization of spin coherence, relaxation, and noise spectra of shallow NV centers near a pristine diamond surface in UHV, together with comparison to measurements under ambient conditions, provides new insight into the microscopic origins of surface-induced decoherence.

Future studies of the temperature dependence of NV spin dynamics and noise spectra will provide further insight into the microscopic mechanisms governing relaxation and decoherence processes. In situ surface modification techniques, such as controlled gas dosing experiments inside the UHV chamber \cite{neethirajan2023controlled}, may provide strategies to compensate for excessive surface charging and engineer more favorable surface environments. Furthermore, investigations of different diamond surface orientations, surface terminations, and diamond heterostructures with other materials will provide complementary insights into the microscopic origins of diamond surface noise.

More broadly, the combination of controlled surface preparation with quantum noise spectroscopy provides a general framework for identifying microscopic loss mechanisms in solid-state quantum systems. This approach can be applied to other qubit platforms, including superconducting qubits, quantum dots, and trapped-ion systems, where surface-induced noise remains a critical limitation for quantum technologies \cite{brown2021materials, chatterjee2021semiconductor, de2021materials, hite2013surface}.

\begin{acknowledgments}
We acknowledge useful discussions with Alastair Stacey and Manik Goyal. This work was primarily supported by the Center for Molecular Quantum Transduction (CMQT), an Energy Frontier Research Center funded by
the U.S. Department of Energy, Office of Science, Basic Energy Sciences under Contract No. DE-SC0021314. Instrumentation and method development was also supported by the U.S. Department of Energy, Office of Science, Office of Basic Energy Sciences, under Award Number DE-SC0018978 and the NSF under the CAREER program (Grant No. DMR- 1752047). Diamond surface characterization was supported by the U.S. Department of Energy, Office of Science,
National Quantum Information Science Research Centers, Codesign Center for Quantum Advantage (C2QA) under Contract No. DESC0012704.  We also acknowledge the support from Princeton Imaging and Analysis Center.
\end{acknowledgments}

\section*{DATA AVAILABILITY}

The data that support the findings of this article are not publicly available. The data are available from the authors upon reasonable request.

\appendix

\section{Diamond sample preparation\label{sec:sample preparation}}

The diamond sample used in this work consists of a $^{12}$C-enriched ($>99.99\%$) as-grown layer on an electronic-grade substrate provided by Element Six. The isotopically purified layer suppresses magnetic noise from $^{13}$C nuclear spins, enabling the use of NV centers in this sample to probe surface-induced decoherence mechanisms.

Shallow NV centers were created by ion implantation of $^{15}$N at an energy of $3$~keV and a dose of $1\times10^9$~cm$^{-2}$. Following implantation, the sample was processed using the procedure described in Ref.~\cite{Sangtawesin2019a}. Prior to loading into the UHV chamber, the diamond sample was cleaned using a piranha solution. The absence of detectable surface contaminants, such as sodium, chlorine, and silicon, could be confirmed by in situ X-ray photoelectron spectroscopy (XPS) within the UHV chamber. Representative XPS spectra are presented in Ref.~\cite{yuan2026integratedultrahighvacuumcluster}.

Prior to UHV characterizations, the sample was annealed in situ to remove surface adsorbates \cite{laikhtman2004interaction,Sangtawesin2019a}. The sample temperature was ramped to $350\,^{\circ}\mathrm{C}$ at a rate of $5\,^{\circ}\mathrm{C/min}$ and maintained at this temperature for $1$~hour. During the annealing process, the UHV chamber pressure remained at approximately $1\times10^{-9}\,\mathrm{mbar}$. This step ensures a well-defined and reproducible surface condition during measurements. In addition, it suppresses laser-induced surface photoluminescence (PL) enhancement associated with laser parking effects (see Ref.~\cite{yuan2026integratedultrahighvacuumcluster} for details), which could otherwise obscure stable optical readout of individual NV centers.

\section{Locating the same NV centers across setups\label{sec:air UHV confocals}}

To enable a direct comparison of NV center properties under different surface conditions, it is essential to measure the same individual NV centers with calibrated depths. Variations in NV depth and local environment can lead to significant differences in spin properties, making such one-to-one correspondence critical for isolating surface-induced effects.

Identifying the same NV centers across different confocal microscope setups presents several experimental challenges. Due to differences in optical configurations, confocal images acquired from different setups can vary in spatial resolution, collection efficiency, field of view, and image orientation. To account for these differences, each setup is calibrated independently using either a standard imaging target or the micrometer readings of the sample translation stage to get an accurate spatial scale factor.
The diamond sample is mounted in each setup with a controlled orientation such that the NV symmetry axes are aligned with the applied external magnetic field. This ensures consistent spin resonance conditions and facilitates comparison of spin properties. After fixing the sample orientation, the lateral confocal scan directions are adjusted so that images with consistent orientations can be obtained across different setups. 

To reliably relocate the same NV centers, we employ a multi-step registration procedure. First, the displacement between the region of interest and a fixed reference point (e.g., a sample corner) is recorded. In addition, large-area confocal scans ($40~\mu$m $\times$ $40~\mu$m) surrounding the NV region are acquired and stored as reference images. When re-imaging the sample in a different setup, we compute the spatial cross-correlation between newly acquired images and the reference scans to determine relative translations and identify the corresponding region. This correlation-based approach enables fast identifications of the same NV batch locations despite differences in imaging conditions.

\begin{figure}
    \centering
    \includegraphics{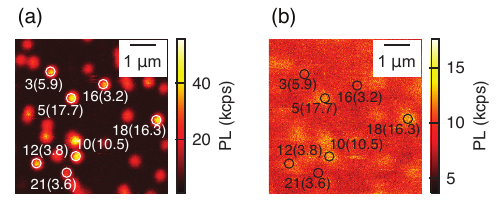}
    \caption{
(a) Confocal scan of a diamond sample acquired under ambient conditions. (b) Confocal scan of the same region acquired under UHV. NV centers are labeled by index, with the measured depth indicated in parentheses (in nm). The excitation laser power is 0.56~mW in air and 1.31~mW in UHV. Several shallow NV centers (e.g., NV3, NV16, and NV21) become undetectable in the UHV scan.}
    \label{fig:confocal_air_UHV}
\end{figure}

A comparison of confocal scans of the same sample region acquired in air and under UHV conditions is shown in Fig.~\ref{fig:confocal_air_UHV}. Relative to the air measurement, the UHV confocal image exhibits reduced spatial resolution of individual NV spots and lower PL count rates. These differences are primarily attributed to the performance limitations of the UHV-compatible imaging objective. Specifically, an Attocube LT-APO/VISIR objective (numerical aperture, NA = 0.82) is used in UHV, while an Olympus MPlanFL N 100x objective (NA = 0.90) is employed for measurements in air. As a consequence of the reduced NV PL collection efficiency and signal-to-noise ratio in UHV, longer averaging times are required for all measurements. We also observe that several of the shallowest NV centers (e.g., NV3, NV16, and NV21) are not visible in the UHV confocal scan. We attribute this absence to charge-state conversion from NV$^{-}$ to NV$^{0}$ for centers located very close to the surface, induced by the removal of surface adsorbates in UHV \cite{neethirajan2023controlled}.

Similarly, we locate the same batch of NV centers in an oil-immersion confocal microscope. The NV depths reported in the paper are calibrated by measuring the proton NMR signal originating from the immersion oil \cite{pham2016nmr}.

\section{NV NMR detection of proton spins from surface adsorbates\label{sec:proton NMR}}

To verify the removal of surface adsorbates and confirm that NV dynamics is measured under adsorbate-free conditions in UHV, we use the NV center as a probe to detect proton NMR signals originating from surface-bound species, which have previously been detected in ambient measurements \cite{xu2025minimizing,loretz2014nanoscale, degen2009nanoscale}.

Under an external magnetic field of $420$~G aligned with the NV axis, proton nuclear spins precess at a Larmor frequency of $1.788$~MHz. This precession can be detected using dynamical decoupling sequences applied to the NV center. When the evolution time of the sequence matches half of the inverse of the target spin precession frequency, the NV coherence exhibits a characteristic dip due to resonant coupling to the proton spins. For the present conditions, this resonance is expected at an evolution time of approximately $280$~ns. Quantum interpolation pulse sequences are used for improved time resolution in this measurement \cite{ajoy2017quantum}.

\begin{figure}
    \centering
    \includegraphics{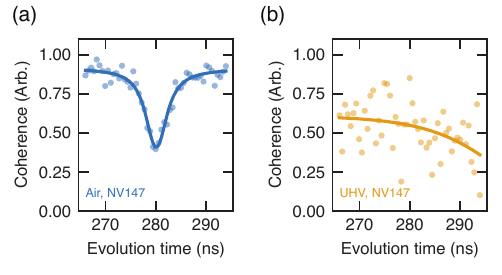}
    \caption{
    Dynamical decoupling measurements using XY8-32 pulse sequences for detection of the proton NMR signal in (a) air and (b) UHV. In both cases, a magnetic field of 420~G is applied along the NV axis. A proton NMR signal is expected at an evolution time of $\sim 280$~ns and is clearly observed in the air measurement. The air data are fitted using an NV NMR response model with a kernel function that accounts for the finite $T_2^*$ of the proton spins~\cite{pham2016nmr}. In contrast, the UHV data exhibit reduced coherence over the entire evolution window, consistent with enhanced noise under UHV conditions, and are fitted by an exponential decay function.
}
    \label{fig:H_NMR}
\end{figure}

Figure~\ref{fig:H_NMR} shows a comparison of dynamical decoupling measurements performed in air and under UHV conditions. In air, a pronounced dip is observed at the expected evolution time, indicating the presence of proton spins from surface adsorbates. In contrast, no such feature is observed in UHV measurements, proving that surface adsorbates have been effectively removed. We checked seven different NV centers in UHV using dynamical decoupling sequences with varying numbers of $\pi$-pulses, and no proton NMR signature was detected.

To quantify the surface proton density, we adapt the model of Ref.~\cite{pham2016nmr} by assuming a two-dimensional distribution of proton nuclear spins on the diamond surface with an areal density $\sigma$. The magnetic field variance produced by this spin bath at an NV center located a depth $d_{\mathrm{NV}}$ beneath the diamond (100) surface is 
\begin{equation}
    B_{\mathrm{RMS}}^2 = \sigma \left(\frac{\mu_0\hbar\gamma_n}{4\pi}\right)^2 \left(\frac{5\pi}{32d_{\mathrm{NV}}^4}\right).
\end{equation}
By fitting the data in Fig.~\ref{fig:H_NMR}, we estimate a surface proton spin density of $221.4\pm23.8~\mathrm{nm}^{-2}$ under ambient conditions. Following UHV annealing, the absence of a detectable proton NMR signal places an upper bound of $23.4~\mathrm{nm}^{-2}$ on the surface proton density.

\section{Measurements on deep NV centers\label{sec:t2 deep nv}}
Figure~1(c) in the main text shows that the ratio $T_{2,\mathrm{air}}/T_{2,\mathrm{UHV}}$ increases for shallower NV centers. To further confirm that this effect is specific to near-surface NV centers, we perform control measurements on deep, naturally occurring NV centers.

These deep NV centers originate from native $^{14}$N impurities in the diamond and are typically located more than $1\,\mu\mathrm{m}$ below the surface. In contrast, the shallow NV centers studied in this work are created by implantation of $^{15}$N ions and reside within approximately 20~nm of the surface. The two types of NV centers can also be distinguished via their characteristic hyperfine splittings.

\begin{figure}
    \centering
    \includegraphics{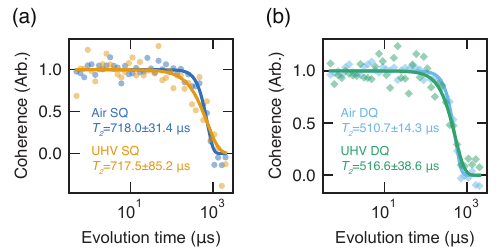}
    \caption{SQ (a) and DQ (b) $T_2$ measurements for deep, naturally occurring NV centers under ambient conditions and in UHV. No measurable difference is observed between air and UHV, indicating that the reduction in $T_2$ arises from surface-related effects. Experimental data (points) are fitted to exponential functions (lines). }
    \label{fig:deep_NV}
\end{figure}

Figure~\ref{fig:deep_NV} shows a comparison of both SQ and DQ Hahn echo $T_2$ measurements for representative deep NV centers in air and under UHV conditions. In contrast to the behavior observed for shallow NV centers, no significant difference in $T_2$ is observed between the two environments for deep NV centers, showing that the observed changes of coherence in UHV originate from modifications to the near-surface noise environment rather than bulk properties of the diamond or technical artifacts.

\section{Effect of green laser excitation\label{sec:t2 green laser}}

\begin{figure}[h]
    \centering
    \includegraphics{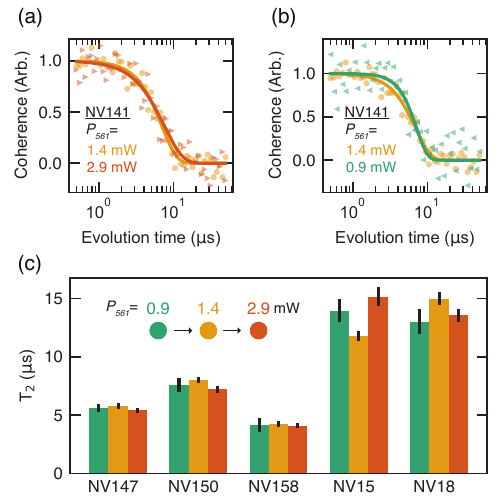}
    \caption{
    (a,b) Representative SQ Hahn echo $T_2$ measurements for a single NV center under UHV conditions at different green excitation powers, showing no measurable change in $T_2$.  Experimental data (points) are fitted to exponential functions (lines). 
    (c) SQ Hahn echo $T_2$ times for five NV centers measured at three different green excitation powers, showing no observable dependence on excitation power.
    }
    \label{fig:green_power}
\end{figure}

Due to differences in the imaging objectives used in the air and UHV confocal setups (see Appendix~\ref{sec:air UHV confocals}), higher green laser power is used for NV measurements in the UHV setup. Specifically, typical excitation powers of $1.4$~mW are used in UHV, compared to $0.5$~mW in air measurements. To evaluate whether this increased optical power influences the observed NV coherence, we measure Hahn echo $T_2$ times in UHV at three different green laser powers: $0.9$~mW, $1.4$~mW, and $2.9$~mW. Across this range, no significant change in $T_2$ is observed (Fig.~\ref{fig:green_power}). In addition, to test for possible transient effects induced by optical excitation, we introduce a $10~\mu$s delay between the end of the green laser pulse and the start of the Hahn echo microwave sequence. This modification likewise produces no measurable change in the extracted $T_2$ values. These results indicate that the enhanced decoherence observed in UHV is not related to the higher optical excitation power used in these measurements.

\section{Effect of ion pump and ion gauge operation\label{sec:ion pump ion gauge}}

During UHV measurements, an ion pump is used to maintain the chamber pressure, and an ion gauge is employed for pressure monitoring. Both devices operate by ionizing residual gas molecules using energetic electrons. This process can, in principle, generate reactive species or radicals \cite{zikovsky2009reaction}. To evaluate whether such effects contribute to the increased electric and magnetic noise in UHV, we perform control experiments with both the ion pump and ion gauge turned off.

To maintain UHV conditions, we replace the ion pump with a turbomolecular pump, which removes gas molecules through mechanical momentum transfer rather than ionization. With the ion pump in operation, the chamber pressure is $5.2\times10^{-10}$~mbar. After switching to the turbomolecular pump, the pressure remains stable at $7.1\times10^{-10}$~mbar, which is measured before turning off the ion gauge.

\begin{figure}
    \centering
    \includegraphics{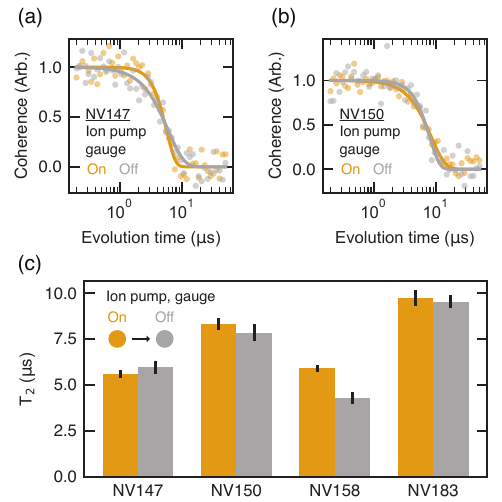}
    \caption{
    (a,b) Representative SQ Hahn echo $T_2$ measurements for two NV centers under UHV conditions with the ion pump and ion gauge turned on and off, showing no measurable change in $T_2$.  Experimental data (points) are fitted to exponential functions (lines). 
    (c) SQ Hahn echo $T_2$ times for four NV centers measured when the ion pump and gauge are on and off, showing no observable dependence on vacuum instrumentation conditions.
    }
    \label{fig:IG_pump}
\end{figure}

Figure~\ref{fig:IG_pump} shows the Hahn echo $T_2$ measurements of NV centers after both the ion pump and the ion gauge are turned off. Within experimental uncertainty, the measured $T_2$ times are unchanged compared to those obtained with the ion pump and ion gauge operating, indicating that the enhanced electric and magnetic noise observed in UHV is not induced by the ionization processes associated with the ion pump or ion gauge.

\section{Decoherence rates decomposition from T2 measurements\label{sec:decoherence rates decomposition}}
To disentangle the contributions of electric and magnetic noise to the observed coherence times, we follow the Lindblad formalism introduced in Ref.~\cite{candido2024interplay}. In this framework, decoherence arises from fluctuations in both electric and magnetic fields, characterized by their respective spectral densities.

The coupling of the NV center to electric field noise gives rise to three rates,
\begin{align}
    \Gamma_{d_\perp}(\omega) &= \tilde{d}^2_\perp \left[S_{E_x}(\omega) + S_{E_y}(\omega) \right], \\
    \Gamma_{d^\prime}(\omega) &= \tilde{d}^{\prime 2} \left[S_{E_x}(\omega) + S_{E_y}(\omega) \right], \\
    \Gamma_{d_\parallel}(\omega) &= \tilde{d}^2_\parallel S_{E_z}(\omega),
\end{align}
while magnetic field noise contributes through
\begin{align}
    \Gamma_{\gamma_\perp}(\omega) &= \tilde{\gamma}^2_\perp \left[S_{B_x}(\omega)+S_{B_y}(\omega)\right], \\
    \Gamma_{\gamma_\parallel}(\omega) &= \tilde{\gamma}^2_\parallel S_{B_z}(\omega).
\end{align}

The decoherence of the off-diagonal density matrix elements in the spin basis follows
\begin{equation}
\frac{d}{dt} \rho_{\mu\nu}(t) = -\frac{1}{T_2^{\mu\nu}} \rho_{\mu\nu}(t),
\end{equation}
with coherence times determined by combinations of the above rates. In the most general form, each $T_2^{\mu\nu}$ contains contributions from both zero-frequency (pure dephasing) noise and finite-frequency noise associated with transitions between spin sublevels:

\begin{align}
\frac{1}{T_2^{0+}} &= \frac{1}{2} [\Gamma_{\gamma_\parallel}(0) + \Gamma_{\gamma d'}(\omega_{+0}) + \Gamma_{d_\perp}(\omega_{+-}) + \Gamma_{d_\parallel}(0)] \nonumber \\
&\quad + \frac{1}{4} \Gamma_{\gamma d'}(\omega_{-0}),
\end{align}
\begin{align}
\frac{1}{T_2^{0-}} &= \frac{1}{2} [\Gamma_{\gamma_\parallel}(0) + \Gamma_{\gamma d'}(\omega_{-0}) + \Gamma_{d_\perp}(\omega_{+-}) + \Gamma_{d_\parallel}(0)] \nonumber \\
&\quad + \frac{1}{4} \Gamma_{\gamma d'}(\omega_{+0}), 
\end{align}
\begin{equation}
\frac{1}{T_2^{-+}} = 2\Gamma_{\gamma_\parallel}(0) + \Gamma_{d_\perp}(\omega_{+-}) + \frac{1}{4} [\Gamma_{\gamma d'}(\omega_{-0}) + \Gamma_{\gamma d'}(\omega_{+0})], 
\end{equation}
where $T_2^{0+}$, $T_2^{0-}$ are SQ coherence times measured between the $\{m_s=0,m_s=+1\}$ and $\{m_s=0,m_s=-1\}$ states, respectively, while $T_2^{-+}$ is the DQ coherence time between the $\{m_s=-1,m_s=+1\}$ states.

These expressions can be simplified when considering the NV measurement conditions in this work. Since the noise spectral densities decrease with increasing frequency, the zero-frequency components dominate in the NV $T_2$ rates,
\begin{equation}
\Gamma_{\gamma_\parallel}(0), \ \Gamma_{d_\parallel}(0) \gg \Gamma_{\gamma d'}(\omega_{\pm 0}), \ \Gamma_{d_\perp}(\omega_{+-}).
\end{equation}
This approximation is independently supported by comparing the NV SQ and DQ $T_1$ and $T_2$ times. The $T_1$ times are determined by finite-frequency transition rates and are typically in the millisecond regime, which is several orders of magnitude longer than the $T_2$ times.
Neglecting the subleading finite-frequency terms, the coherence times reduce to
\begin{align}
\frac{1}{T_2^{0+}} &= \frac{1}{2} \left[\Gamma_{\gamma_\parallel}(0) + \Gamma_{d_\parallel}(0)\right], \\
\frac{1}{T_2^{0-}} &= \frac{1}{2} \left[\Gamma_{\gamma_\parallel}(0) + \Gamma_{d_\parallel}(0)\right], \\
\frac{1}{T_2^{-+}} &= 2\,\Gamma_{\gamma_\parallel}(0).
\end{align}

These expressions provide a direct route to decomposing the total dephasing rate into magnetic and electric contributions. In particular, the $(-,+)$ coherence is insensitive to electric field noise at leading order and depends only on longitudinal magnetic fluctuations. This allows us to extract
\begin{equation}
\Gamma_{\gamma_\parallel}(0) = \frac{1}{2T_2^{-+}}.
\end{equation}
Substituting this into the expressions for $T_2^{0\pm}$, we isolate the electric contribution,
\begin{equation}
\Gamma_{d_\parallel}(0) = \frac{2}{T_2^{0\pm}} - \frac{1}{2T_2^{-+}}.
\end{equation}

Following the notation in the main text, we define the magnetic and electric dephasing rates as
\begin{align}
\frac{1}{T_{2,B}} &= \frac{1}{2}\Gamma_{\gamma_\parallel}(0), \\
\frac{1}{T_{2,E}} &= \frac{1}{2}\Gamma_{d_\parallel}(0),
\end{align}
such that the total dephasing rate in the $(0,\pm)$ subspaces can be written as a simple sum,
\begin{equation}
\frac{1}{T_2^{0\pm}} = \frac{1}{T_{2,B}} + \frac{1}{T_{2,E}},
\end{equation}
which matches the decomposition shown in Eq.~\ref{eq:T2_SQ} in Sec.~\ref{sec:E_B_T2}.

We note that this expression is derived within the Lindblad framework assuming Markovian noise dynamics. While it provides a useful phenomenological decomposition of the magnetic and electric contributions, deviations may arise in experiments due to finite noise correlation times.

\section{Modified surface charge model and depth scaling of noise\label{sec:depth scaling model}}

From the fits in Fig.~\ref{fig:DQ_T2}(f), we extract depth-scaling exponents smaller than 2 for the air measurements. This deviation from the expected $1/d_{\mathrm{NV}}^{2}$ behavior can be understood by extending the surface charge model in Ref.~\cite{candido2024interplay} to account for a finite spatial distribution of charges above the diamond surface.

For fluctuating point-like charges confined to an ideal two-dimensional (2D) surface, the electric field correlation functions at the NV location are given by
\begin{equation}
\langle E_x^p(t)\, E_x^p(0) \rangle
= \left( \frac{e}{4\pi \epsilon} \right)^2 
\frac{\pi n_S}{4 d_{\mathrm{NV}}^2} \, f(t),
\end{equation}
\begin{equation}
\langle E_y^p(t)\, E_y^p(0) \rangle
= \left( \frac{e}{4\pi \epsilon} \right)^2 
\frac{\pi n_S}{8 d_{\mathrm{NV}}^2} \, (3-\cos 2\theta)\, f(t),
\end{equation}
\begin{equation}
\langle E_z^p(t)\, E_z^p(0) \rangle
= \left( \frac{e}{4\pi \epsilon} \right)^2 
\frac{\pi n_S}{8 d_{\mathrm{NV}}^2} \, (3+\cos 2\theta)\, f(t),
\end{equation}
where $n_S$ is the surface charge density, $d_{\mathrm{NV}}$ is the NV depth, $\theta$ is the angle between the NV axis and the surface normal, and $f(t)$ is the temporal correlation function.

To capture the effect of adsorbates in ambient conditions, we instead consider a finite slab of thickness $D$ with a volumetric charge density $\eta_V$. In this case, the electric field correlations become
\begin{equation}
\langle E_x^p(t)\, E_x^p(0) \rangle
= \left( \frac{e}{4\pi \epsilon} \right)^2 
\frac{\pi \eta_V D}{4 d_{\mathrm{NV}} (d_{\mathrm{NV}} + D)} \, f(t),
\end{equation}
\begin{equation}
\langle E_y^p(t)\, E_y^p(0) \rangle
= \left( \frac{e}{4\pi \epsilon} \right)^2 
\frac{\pi \eta_V D}{8 d_{\mathrm{NV}} (d_{\mathrm{NV}} + D)} \, (3-\cos 2\theta)\, f(t),
\end{equation}
\begin{equation}
\langle E_z^p(t)\, E_z^p(0) \rangle
= \left( \frac{e}{4\pi \epsilon} \right)^2 
\frac{\pi \eta_V D}{8 d_{\mathrm{NV}} (d_{\mathrm{NV}} + D)} \, (3+\cos 2\theta)\, f(t).
\end{equation}
In this model, the depth dependence of the electric noise is modified from $1/d_{\mathrm{NV}}^{2}$ to $D/[d_{\mathrm{NV}}(d_{\mathrm{NV}} + D)]$. A similar modification applies to magnetic noise arising from the motion of charged particles.

Within this framework, the reduced depth-scaling exponent observed in air can be attributed to the presence of an adsorbate layer that redistributes surface charges over a finite thickness, effectively smoothing the spatial dependence of the noise. In contrast, under UHV conditions, the absence of adsorbates confines charges closer to an ideal 2D surface, restoring a scaling closer to $1/d_{\mathrm{NV}}^{2}$.

We note that distributing charges within a finite slab increases the effective separation between the NV center and the fluctuating charges, which would generally reduce the noise amplitude. However, quantitatively accounting for the observed differences between air and UHV would require a slab thickness $D \gtrsim 3 d_{\mathrm{NV}}$. Given that the NV centers studied here are typically deeper than $\sim$10~nm, this would imply an adsorbate layer thickness exceeding tens of nanometers, which is not physically reasonable.

We therefore conclude that while charge redistribution within a finite layer can qualitatively explain the modified depth scaling in air, it cannot fully account for the observed noise differences. Instead, the enhanced noise in UHV is more likely dominated by an increased surface charge density and altered charge dynamics.

\section{$1/f^{\alpha}$ noise spectrum\label{sec:1/f spec}} 

The $1/f^{\alpha}$ noise spectrum extracted from the NV spectral decomposition measurements in Fig.~\ref{fig:DD_spec} can be understood as arising from an ensemble of Lorentzian fluctuators. A single fluctuator with characteristic correlation time $\tau$, such as electron trapping and detrapping at charge defects or charge diffusion processes, produces a Lorentzian noise spectrum of the form
\begin{equation}
    S(\omega)\propto \frac{\tau}{1+\omega^2\tau^2}.
\end{equation}

For near-surface NV centers, the environment is expected to contain a broad distribution of fluctuators with different correlation times $\tau$. We model this distribution as
\begin{equation}
    p(\tau)=\frac{A}{\tau^{2-\alpha}},
\end{equation}
where $A$ is a normalization constant. The resulting total noise spectrum is then
\begin{equation}\label{eq:Spectral_integral}
    S(\omega)\propto \int_{\tau_1}^{\tau_2}\frac{\tau}{1+\omega^2\tau^2}p(\tau)\,d\tau,
\end{equation}
where $\tau_1$ and $\tau_2$ represent the lower and upper cutoff correlation times, respectively. From the experimentally observed spectra, we estimate $\tau_1<10^{-6}$~s and $\tau_2>10^{-4}$~s, allowing the intermediate-frequency regime probed by NV spectral decomposition to exhibit a $1/f^{\alpha}$ dependence.

We now derive the exact solution to the integral in Eq.~\ref{eq:Spectral_integral} which can be related to the Euler Beta functions, beginning with the coordinate transformation $x=\omega \tau$:
\begin{equation}
    S(\omega)\propto\frac{A}{\omega^{\alpha}}\int_{x_{1}\equiv\omega \tau_{1}}^{x_{2}\equiv\omega\tau_{2}}\frac{x^{\alpha-1}}{1+x^{2}}\,dx
\end{equation}
This is followed by two subsequent coordinate transformations, $u=x^{2}$ and $T=u/(1+u)$. The result is:
\begin{align}
    S(\omega)&\propto\frac{A}{2\omega^{\alpha}}\int_{T_{1}\equiv T(u(x_{1}))}^{T_{2}\equiv T(u(x_{2}))}\left(1-T\right)^{-\frac{\alpha}{2}}T^{\frac{\alpha}{2}-1}\,dT\notag\\
    &=\frac{A}{2\omega^{\alpha}}\left(\int_{0}^{T_{2}}f_{\alpha}(T)\,dT-\int_{0}^{T_{1}}f_{\alpha}(T)\,dT\right)\label{eq:pre_Beta_int}
\end{align}
where $f_{\alpha}(T)=\left(1-T\right)^{-\frac{\alpha}{2}}T^{\frac{\alpha}{2}-1}$. The integrals in Eq.~\ref{eq:pre_Beta_int} map directly to the incomplete Beta function, $B(z;a,b)\equiv\int_{0}^{z}(1-T)^{b-1}T^{a-1}\,dT$, as follows:
\begin{align}
    S(\omega)\propto\frac{A}{2\omega^{\alpha}}\biggr[B\left(T_{2};\frac{\alpha}{2},1-\frac{\alpha}{2}\right)-B\left(T_{1};\frac{\alpha}{2},1-\frac{\alpha}{2}\right)\biggr]
\end{align}
Finally, using the relationships between the incomplete Beta function, Euler Beta function, and Euler Gamma function along with the Euler reflection formula; the final result is:
\begin{align}\label{eq:spectral_Beta_result}
    S(\omega)\propto\frac{A}{\omega^{\alpha}}&\frac{\pi}{2\sin(\frac{\pi\alpha}{2})}\notag\\-\frac{A}{2\omega^{\alpha}}&\biggr[B\left(1-T_{2};1-\frac{\alpha}{2},\frac{\alpha}{2}\right)+B\left(T_{1};\frac{\alpha}{2},1-\frac{\alpha}{2}\right)\biggr]
\end{align}
We note that the second and third terms in Eq.~\ref{eq:spectral_Beta_result} vanish as $\tau_{1}\rightarrow0$ and $\tau_{2}\rightarrow\infty$. Thus, the first term is the intermediate regime of $S(\omega)\propto1/\omega^{\alpha}$, while the second and third terms are the corrections for finite integration bounds of $\tau_{2}$ and $\tau_{1}$, respectively.

We now consider three frequency regimes.

\textit{(1) Low-frequency regime: $\omega \ll 1/\tau_2$.} In this limit,
\begin{equation}
    S(\omega)\propto \int_{\tau_1}^{\tau_2}\tau p(\tau)\,d\tau
    =A\frac{\tau_2^\alpha-\tau_1^\alpha}{\alpha},
\end{equation}
which is independent of $\omega$, corresponding to white noise.

\textit{(2) High-frequency regime: $\omega \gg 1/\tau_1$.} In this case,
\begin{equation}
    S(\omega)\propto \int_{\tau_1}^{\tau_2}\frac{1}{\omega^2\tau}p(\tau)\,d\tau
    =A\frac{\tau_1^{\alpha-2}-\tau_2^{\alpha-2}}{2-\alpha}\frac{1}{\omega^2},
\end{equation}
which yields the characteristic $1/\omega^2$ scaling.

\textit{(3) Intermediate-frequency regime: $1/\tau_2 \ll \omega \ll 1/\tau_1$.} In this regime,
\begin{equation}
    S(\omega)\propto \int_{\tau_1}^{\tau_2}\frac{\tau}{1+\omega^2\tau^2}p(\tau)\,d\tau
    =\frac{A}{\omega^\alpha}\int_{\omega\tau_1}^{\omega\tau_2}\frac{x^{\alpha-1}}{1+x^2}\,dx,
\end{equation}
where we have substituted $x=\omega\tau$. Since $\omega\tau_1\ll1$ and $\omega\tau_2\gg1$ in this regime, the integration limits can be extended to $(0,\infty)$, giving
\begin{equation}
    S(\omega)\propto \frac{A}{\omega^\alpha}\int_0^\infty\frac{x^{\alpha-1}}{1+x^2}\,dx
    =\frac{A}{\omega^\alpha}\frac{\pi}{2\sin\left(\frac{\pi\alpha}{2}\right)}.
\end{equation}
This reproduces the experimentally observed $1/\omega^\alpha$ dependence.

Using this model, we can further estimate the surface charge density from the measured NV noise spectral density. Following the fluctuating surface charge model developed in Ref.~\cite{candido2024interplay}, the electric-field noise spectrum along the NV axis is given by
\begin{align}
    S_{E_z^p}(\omega)
    =
    \left(\frac{e}{4\pi\epsilon}\right)^2
    \frac{\pi n_S}{8d_{\mathrm{NV}}^2}
    (3+\cos2\theta)
    \int_{\tau_1}^{\tau_2}
    \frac{2\tau_p}{1+\omega^2\tau_p^2}
    p(\tau_p)\,d\tau_p,
\end{align}
where $d_{\mathrm{NV}}$ is the NV depth, $n_S$ is the surface charge density, and $\theta$ is the angle between the NV axis and the diamond surface normal. For the (100)-oriented diamond surface used in this work, $\theta=54.7^\circ$.

Assuming the same power-law distribution of fluctuators,
\begin{equation}
    p(\tau_p)=\frac{A}{\tau_p^{2-\alpha}},
\end{equation}
the intermediate-frequency regime ($1/\tau_2 \ll \omega \ll 1/\tau_1$) yields
\begin{align}
    S_{E_z^p}(\omega)
    =
    \left(\frac{e}{4\pi\epsilon}\right)^2
    \frac{\pi n_S}{8d_{\mathrm{NV}}^2}
    (3+\cos2\theta)
    \frac{2A}{\omega^\alpha}
    \frac{\pi}{2\sin\left(\frac{\pi\alpha}{2}\right)}.
\end{align}
By comparing this expression with the fitted spectra in Fig.~\ref{fig:DD_spec}, we estimate the corresponding surface charge density. The total charge density $n_S$ depends on the normalization constant $A$, which in turn depends on the cutoff times $\tau_1$ and $\tau_2$. However, the distribution density $n_Sp(\tau_p)$ within the experimentally relevant correlation-time range is independent of the specific choice of $A$. Therefore, we estimate the density of fluctuating charges with correlation times in the range $10^{-6}<\tau_p<10^{-4}$~s, corresponding to the frequency window probed in the NV measurements.

For NV5, the estimated surface charge density in air is $(8.4\pm3.7)\times10^{14}~\mathrm{m}^{-2}$, while under UHV conditions the estimated density increases to $(6.2\pm4.9)\times10^{15}~\mathrm{m}^{-2}$. In this analysis, the dielectric constant $\epsilon$ is taken to be the average of the dielectric constants of air and diamond. We further note that the measured spectra in Fig.~\ref{fig:DD_spec} include contributions from both magnetic and electric noise sources. To obtain an order-of-magnitude estimate of the surface charge density, here we assume that 50\% of the measured noise arises from axial electric-field fluctuations generated by surface charge dynamics, consistent with the analysis in Fig.~\ref{fig:DQ_T2}. When comparing the relative strengths of electric and magnetic noise using the models in Ref.~\cite{candido2024interplay}, we note that the surface density of charges responsible for electric noise may differ from the density of charges whose motion generates magnetic noise. The correlation time distributions associated with trapping/release processes of charges responsible for electric noise and the surface charge dynamics that generate magnetic noise are also expected to be different. However, this uncertainty does not change our estimate that the overall surface charge density is increased in UHV compared to that in air. A more rigorous separation of electric and magnetic noise contributions to the noise spectrum could be achieved through combining dynamical decoupling measurements performed in both the SQ and DQ bases.

\section{Additional $T_1$ data from air and oil confocal setups\label{sec:extra_T1}}
\begin{figure}
    \centering
    \includegraphics{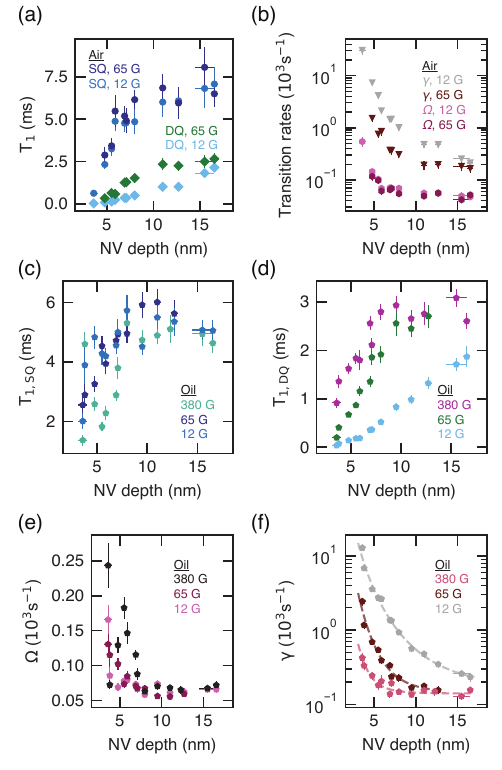}
    \caption{
    (a) Additional SQ and DQ $T_1$ measurements in air at two magnetic fields (12~G and 65~G). 
    (b) Extracted SQ ($\Omega$) and DQ ($\gamma$) relaxation rates in air at the corresponding fields. 
    (c,d) SQ (c) and DQ (d) $T_1$ measurements for NV centers in the same sample acquired using an oil-immersion confocal setup at three magnetic fields: 12~G, 65~G, and 380~G. 
    (e,f) SQ relaxation rate $\Omega$ (e) and DQ relaxation rate $\gamma$ (f) measured in the oil setup as a function of NV depth for the three magnetic fields. The $\gamma$ rates are fitted to a power-law function with an offset, yielding exponents of $-4.06 \pm 1.13$ (380~G), $-3.74 \pm 0.52$ (65~G), and $-3.09 \pm 0.18$ (12~G), and offsets of $0.14 \pm 0.01$, $0.14 \pm 0.02$, and $0.15 \pm 0.03$, respectively.
    }
    \label{fig:extra_air_oil_T1}
\end{figure}

We present additional $T_1$ measurements performed on the same diamond sample in air and in an oil-immersion confocal microscope, as shown in Fig.~\ref{fig:extra_air_oil_T1}. The oil-immersion measurements are carried out using a Nikon Plan Fluor 100$\times$ objective (NA = 1.3) with Nikon type NF immersion oil, and a 520~nm excitation laser.

Both air and oil datasets extend to shallower NV centers than those discussed in the main text. In contrast, spin measurements on NV centers shallower than $\sim$9~nm are not stable under UHV conditions, likely due to charge state depletion of near-surface NV$^-$ centers.

We perform SQ and DQ $T_1$ measurements on NV centers with calibrated depths and extract the corresponding relaxation rates $\Omega$ and $\gamma$. The depth dependence observed in $T_1$ measurements differs from that of $T_2$ shown in Fig.~\ref{fig:DQ_T2}. For the SQ relaxation rate $\Omega$, NV centers shallower than $\sim$7~nm exhibit a stronger depth dependence, while for deeper NV centers---such as those presented in Fig.~\ref{fig:T1} of the main text---the dependence becomes less pronounced.

For the DQ relaxation rate $\gamma$, we observe a clear dependence on the applied magnetic field (i.e., transition frequency). Measurements performed at different magnetic fields show longer DQ $T_1$ times and correspondingly lower $\gamma$ rates at higher fields, in agreement with previous reports~\cite{myers2017double,Sangtawesin2019a}. 

In addition, the depth dependence of $\gamma$ exhibits a different scaling behavior from the $T_2$ decoherence rates, we fit the oil-immersion $\gamma$ data as a function of NV depth to a power-law model with an offset (Fig.~\ref{fig:extra_air_oil_T1}(f)). The extracted depth dependence exponents range from 3.09 to 4.06, suggesting a surface noise model more consistent with electric noise arising from fluctuating surface dipoles. The offset captures a depth-independent bulk contribution to the relaxation rate. From these fits, we obtain a consistent offset of approximately $0.14 \times 10^{3}$~s$^{-1}$. Notably, the $\gamma$ values measured under UHV conditions (Fig.~\ref{fig:T1}(f) in the main text) are close to this bulk limit, suggesting that the dominant surface-related contribution to $\gamma$ is suppressed in UHV.

Finally, we do not observe a significant difference between air and oil-immersion measurements, indicating that the presence of the immersion oil does not significantly alter noise sources compared to those under ambient conditions.

\section{NV noise spectrum from $T_2$ and $T_1$ measurements\label{sec:combined spectrum}}

To directly compare the NV noise spectral densities extracted from decoherence ($T_2$) and relaxation ($T_1$) measurements, we express both in terms of an effective transverse electric field spectral density $S_{E_{\perp}}(\omega)$~\cite{Sangtawesin2019a,myers2017double}. This unified representation allows for a consistent comparison across different frequency ranges. The conversion is given by
\begin{equation}
S_{E_{\perp}}^{\mathrm{SQ}}(\omega) = 2 S_{E_{\parallel}}^{\mathrm{SQ}}(\omega) 
= 2\frac{S_{\mathrm{SQ}}(\omega)}{d_\parallel^2}, 
\quad
S_{E_{\perp}}^{\mathrm{DQ}}(\omega) = \frac{S_{\mathrm{DQ}}(\omega)}{d_\perp^2},
\end{equation}
where $d_\parallel = 3.5~\mathrm{mHz\,m/V}$ and $d_\perp = 170~\mathrm{mHz\,m/V}$ are the longitudinal and transverse electric dipole moments of the NV center, respectively.

\begin{figure}
    \centering
    \includegraphics{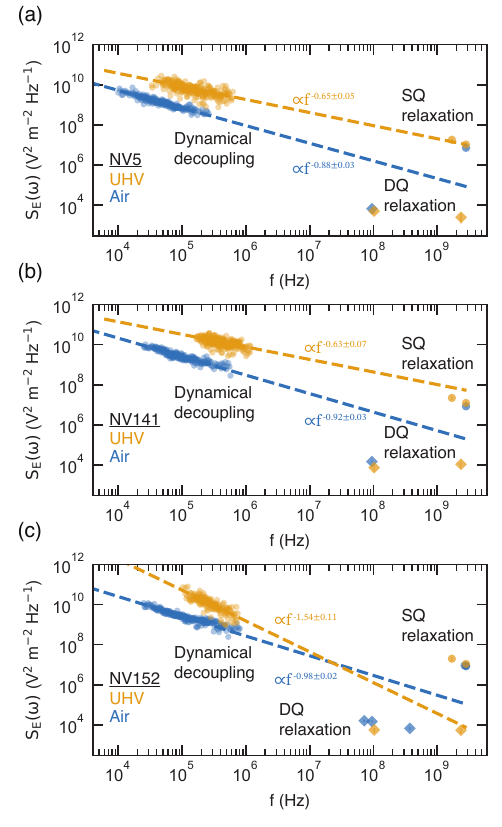}
    \caption{
    Combined noise spectra in terms of transverse electric field power for three NV centers: NV5 (a), NV141 (b), and NV152 (c), measured in air and under UHV conditions. The spectra are constructed by combining dynamical decoupling measurements (low-frequency range) with relaxation spectroscopy (high-frequency range). Dashed lines show power-law fits to the low-frequency data, extrapolated to higher frequencies to allow direct comparison across the two frequency regimes.
    }
    \label{fig:full_spectrum}
\end{figure}

Using this conversion, we combine the low-frequency noise spectra obtained from dynamical decoupling measurements with the high-frequency noise extracted from $T_1$ relaxation spectroscopy. The resulting broadband spectra for three representative NV centers, measured in both air and UHV, are shown in Fig.~\ref{fig:full_spectrum}.

To facilitate comparison, the power-law fits obtained from the dynamical decoupling spectra are extrapolated to the SQ and DQ transition frequencies corresponding to the $T_1$ measurements. We find no clear correlation between the changes in noise observed under UHV conditions at low frequencies (probed by dynamical decoupling) and those at higher frequencies (probed by $T_1$ relaxation). Furthermore, the discrepancy between the extrapolated noise at the SQ and DQ relaxation frequencies---derived from spectral decomposition---and the relaxation rates obtained from SQ and DQ $T_1$ measurements suggests a change in the spectral slope between these two frequency regimes, implying that the upper cutoff frequency of the noise model described in Appendix~\ref{sec:1/f spec} lies between these two regimes.

This lack of correlation suggests that distinct physical mechanisms dominate the noise in different frequency regimes. In particular, the enhanced low-frequency noise observed in UHV likely arises from slow surface charge fluctuation and movement, whereas the high-frequency noise probed by DQ $T_1$ measurements is related to the dynamics of surface adsorbates. These results further support the picture of a frequency-dependent noise environment at the diamond surface.

\section{Long term UHV measurements\label{sec:long term measurements}}

To evaluate the long-term stability of the NV measurements in UHV, we monitor the same diamond sample over a period of several months. Over time, we observe a slight increase in the background PL surrounding the measured NV centers. This increase is more pronounced in cases where the sample is introduced into the UHV chamber without an in situ annealing step to remove surface adsorbates (see Ref.~\cite{yuan2026integratedultrahighvacuumcluster}). This observation could suggest a gradual re-accumulation of surface adsorbates during extended operation in UHV.

\begin{figure}[h]
    \centering
    \includegraphics{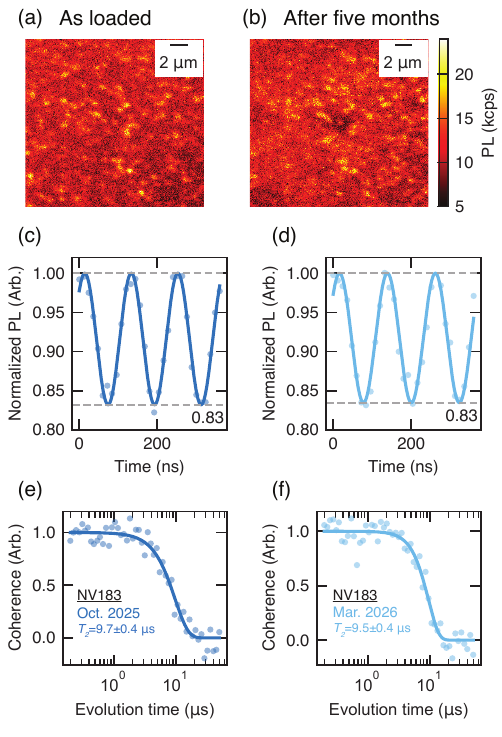}
    \caption{
    (a) UHV confocal scan of the diamond surface around NV183. 
    (b) UHV confocal scan of the same region after five months, showing increased background PL around the central NV; at the NV position, the focused green laser used during measurements quenches the background PL, resulting in a dark central spot. 
    (c,d) Rabi oscillation measurements of NV183 at two different times, showing no observable change in Rabi contrast. Experimental data (points) are fitted to sinusoidal functions (solid lines). The dashed lines indicate the Rabi contrasts.
    (e,f) SQ Hahn echo $T_2$ measurements of NV183 at two different times, showing no observable change in $T_2$. Experimental data (points) are fitted to exponential functions (lines). 
    }
    \label{fig:long_time_surface_PL}
\end{figure}

Despite the increased background PL in the surrounding area, no noticeable increase is observed at the location of the measured NV center itself, where the excitation laser is parked during measurements. As a result, the signal-to-noise ratio of the NV readout remains unchanged. This is confirmed by comparing Rabi oscillation contrasts of the same NV center measured after five months in UHV, which show no measurable degradation (Fig.~\ref{fig:long_time_surface_PL}(c,d)). Furthermore, we do not observe any systematic change in the NV spin coherence times ($T_2$) over this period (Fig.~\ref{fig:long_time_surface_PL}(e,f)).

\section{NV ensemble sample characterization in air and UHV\label{sec:ensemble measurements}}
In addition to single NV measurements, we characterize a shallow NV ensemble sample under both ambient and UHV conditions. The NV ensemble sample is created by high-dose nitrogen ion implantation ($2.5$~keV, $2\times10^{12}$~cm$^{-2}$) into a diamond substrate with a $\sim 100~\mu$m-thick, $^{12}$C-enriched ($99.999\%$) as-grown layer provided by Element Six. All measurements are performed using the same confocal microscope setups described in the main text for air and UHV experiments.

\begin{figure}[h]
    \centering
    \includegraphics{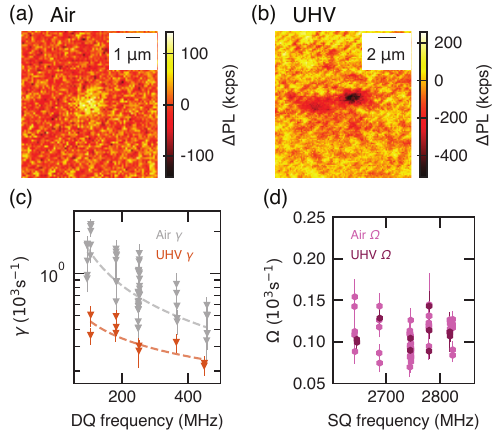}
    \caption{
    (a,b) Difference maps of confocal scans acquired before and after NV $T_1$ measurements at the center spot in air (a) and in UHV (b). In air, an increase in NV ensemble PL is observed at the laser parking position, whereas in UHV a reduction in PL is observed at the measurement location. A slight lateral drift of the sample in UHV results in a dark tail extending to the left of the central spot. 
    (c,d) DQ $\gamma$ relaxation rates (c) and SQ $\Omega$ relaxation rates (d) extracted from ensemble $T_1$ measurements in air and in UHV as a function of frequency. The $\gamma$ rates exhibit a clear frequency dependence and are fitted to power-law functions, yielding exponents of $0.83\pm0.09$ in air and $0.43\pm0.11$ in UHV.
    }
    \label{fig:ensemble}
\end{figure}

We first examine the optical response of the NV ensemble under continuous green laser excitation. In ambient conditions, we observe an increase in NV PL from 570~kcps to 600~kcps in less than an hour at the measurement location where the laser is focused. Green 561~nm laser power is 0.15~mW. In contrast, under UHV conditions, the PL at the measured location typically decreases from 600~kcps to 450~kcps after several minutes of illumination. 561~nm laser power used in UHV is 0.85~mW. Figures~\ref{fig:ensemble}(a,b) show difference maps of the confocal scans before and after measurements in air and UHV, respectively. 

The observed PL increase in air suggests an enhancement of the NV$^{-}$ charge state population under laser exposure, whereas the PL decrease in UHV indicates a shift toward the neutral NV$^{0}$ charge state. Consistent with the single NV results discussed in Sec.~\ref{sec:air UHV confocals}, these observations suggest that NV$^{-}$ centers very close to the surface are depleted under UHV conditions, while in air, green laser excitation promotes stabilization of the NV$^{-}$ charge state.

To probe the noise environment, we perform SQ and DQ $T_1$ measurements on the NV ensemble as well and extract the corresponding relaxation rates $\Omega$ (SQ) and $\gamma$ (DQ). By varying the external magnetic field, we tune the transition frequencies and map out the frequency dependence of the relaxation rates, as shown in Fig.~\ref{fig:ensemble}(c,d). For each magnetic field, measurements are performed at multiple locations across the sample, resulting in a distribution of relaxation rates.

We find that the $\gamma$ relaxation rates measured in air are consistently higher than those in UHV across all sampled locations (Fig.~\ref{fig:ensemble}(c)), in agreement with the trends observed in single NV measurements. However, for the ensemble data, it is less straightforward to attribute this reduction solely to the removal of surface-related noise sources. The observed charge-state depletion in UHV effectively biases the detected signal toward deeper NV centers, which may intrinsically exhibit lower sensitivity to surface noise. On the other hand, no clear difference of the $\Omega$ relaxation rates is observed between air and UHV measurements (Fig.~\ref{fig:ensemble}(d)).

Therefore, while the ensemble measurements qualitatively support the single NV results, they also highlight the importance of depth-calibrated single NV characterizations when interpreting surface noise change under different surface conditions.

\section{Air characterizations directly after unloading from UHV\label{sec:air after UHV}}

\begin{figure*}
    \centering
    \includegraphics{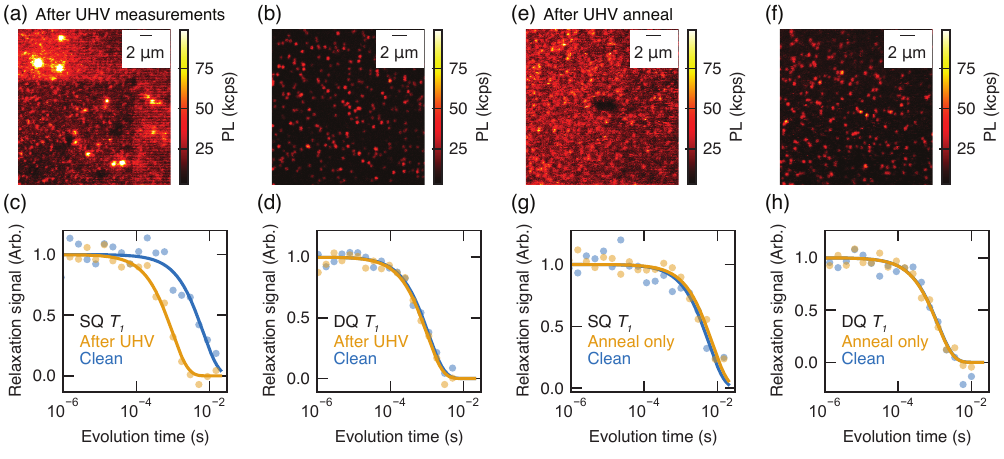}
    \caption{
    (a) Confocal scan of the diamond surface in air immediately after unloading from the UHV chamber without any cleaning. Bright spots correspond to NV centers that are measured for extended periods inside the UHV chamber. Elevated surface PL is observed across the sample surface, and can be suppressed by green laser illumination, as indicated by the darker scanned region in the lower-left portion of the image. 
    (b) Confocal scan of the same region shown in (a) after repeated laser scans and NV measurements. The enhanced PL at the NV locations and the elevated background PL are both reduced to levels typical of a cleaned diamond surface measured in the air confocal setup.
    (c,d) SQ (c) and DQ (d) $T_1$ measurements of NV141 performed immediately after unloading from the UHV chamber (orange) and after piranha cleaning (blue). A substantially shorter $T_{1, \mathrm{SQ}}$ is observed directly after unloading from UHV ($1.06\pm0.10$~ms) compared to the cleaned surface condition ($6.49\pm0.94$~ms). In contrast, no significant change is observed in the DQ relaxation time: $T_{1, \mathrm{DQ}} = 1.12\pm0.12$~ms after unloading from UHV and $1.10\pm0.09$~ms after piranha cleaning. 
    (e) Confocal scan of the diamond surface in air after unloading from an annealing process performed inside the UHV chamber. Elevated background PL similar to that observed in (a) is present. 
    (f) Confocal scan of a different region of the same diamond surface acquired three days after the measurement in (e). The laser power for all confocal scans is approximately 0.56 mW.
    (g,h) SQ (g) and DQ (h) $T_1$ measurements of NV141 performed immediately after unloading from an annealing in UHV (orange) and after cleaning (blue). No significant change is observed in both SQ and DQ relaxation times: $T_{1, \mathrm{SQ}} = 7.13\pm0.76$~ms, $T_{1, \mathrm{DQ}} = 1.18\pm0.12$~ms after unloading from UHV and $T_{1, \mathrm{SQ}} = 5.76\pm0.84$~ms, $T_{1, \mathrm{DQ}} = 1.20\pm0.17$~ms after cleaning. The magnetic field during all $T_1$ measurements is approximately 18 G. Points represent experimental data and solid lines are exponential fits.
    }
    \label{fig:air_directly_after_UHV}
\end{figure*}

After annealing the sample and characterizing individual NV centers inside the UHV chamber, the sample is unloaded to the air confocal setup without any cleaning. We observe elevated surface background PL across the diamond surface, with an increase of approximately 10~kcps under 561~nm laser illumination at 0.56~mW. This enhanced PL can be suppressed by green laser illumination. In the surface region previously characterized inside the UHV chamber, we additionally observe strongly enhanced PL ($>100$~kcps) localized at the positions of NV centers that have undergone extended measurements under UHV conditions (Fig.~\ref{fig:air_directly_after_UHV}(a)). Both the localized PL enhancement at the NV spots and the elevated background PL can be suppressed within less than 1~min under 0.56~mW green laser illumination. After repeated confocal scans and measurements on the NV centers, the enhanced surface PL is completely suppressed to levels comparable to those observed for a cleaned diamond surface measured in the air confocal setup (Fig.~\ref{fig:air_directly_after_UHV}(b)).

As shown in Fig.~\ref{fig:shorter_T2}(d), the NV \(T_2\) values measured immediately after unloading from UHV are comparable to those obtained under a piranha-cleaned surface condition. In contrast, we frequently observe a reduction in SQ \(T_1\), particularly for NV centers located closer to the surface (Fig.~\ref{fig:air_directly_after_UHV}(c)). No obvious change in DQ \(T_1\) is observed (Fig.~\ref{fig:air_directly_after_UHV}(d)).

To determine whether the increased surface PL and shortened SQ \(T_1\) originate from the UHV annealing process itself or from prolonged green laser illumination under UHV conditions, we perform a control experiment in which the sample is unloaded immediately after UHV annealing without any UHV confocal characterization or green laser exposure. After transferring the sample to the air confocal setup without cleaning, we again observe elevated surface PL across the sample surface (Fig.~\ref{fig:air_directly_after_UHV}(e)). Similar to the previous observations, this increased PL can be suppressed by green laser illumination. After keeping the sample in air for three days, we measure a previously uncharacterized region of the sample surface and observe substantially reduced background PL (Fig.~\ref{fig:air_directly_after_UHV}(f)), comparable to that of a cleaned diamond surface. These observations suggest that the enhanced surface PL induced after unloading from UHV can gradually dissipate through prolonged exposure to air.

Notably, after this control UHV annealing experiment, we do not observe enhanced PL localized at individual NV spots, since no extended measurements have been performed under UHV conditions. Furthermore, no obvious reduction in NV SQ \(T_1\) and DQ \(T_1\) is measured (Fig.~\ref{fig:air_directly_after_UHV}(g, h)). These results suggest that the shortening of NV SQ \(T_1\) is associated with prolonged laser exposure under UHV rather than the annealing process itself.

The microscopic origin of the increased surface PL and the change in NV SQ \(T_1\) remains unclear; however, we propose one possible explanation. The observation of enhanced PL across the entire sample surface suggests a global effect associated with UHV annealing. One possibility is that the annealing process induces surface charging under UHV conditions, which gradually dissipates upon exposure to air. During NV measurements in UHV, repeated green laser illumination drives NV charge-state cycling and may generate additional free carriers. These carriers can become trapped near the illuminated NV regions and modify the local surface environment. Upon subsequent exposure to air, these modified regions may interact with hydrocarbons or other adsorbates, leading to the strongly enhanced PL observed at the NV locations. These adsorbates could additionally introduce high-frequency magnetic noise that shortens the SQ \(T_1\). The absence of observable changes in \(T_2\) and DQ \(T_1\) suggests that no significant increase in lower-frequency magnetic noise or electric-field noise is introduced.

\bibliography{apssamp}

\end{document}